\documentclass[aps,preprintnumbers,showpacs,nofootinbib,a4paper]{revtex4}
\usepackage{subfig}
\usepackage{graphicx}
\usepackage{dcolumn}
\usepackage{epsfig}
\usepackage{bm}
\usepackage{amsmath}
\usepackage{amsfonts}
\usepackage{amssymb}
\usepackage{geometry}
\usepackage{epstopdf}
\usepackage{upgreek}
\newcommand{\eq}{\begin{eqnarray}}
\newcommand{\en}{\end{eqnarray}}

\newcommand{\ba}[1]{\begin{eqnarray} \label{(#1)}}
\newcommand{\ea}{\end{eqnarray}}

\newcommand{\newc}{\newcommand}
\newc{\lra}{\leftrightarrow}
\newc{\beq}{\begin{equation}}
\newc{\eeq}{\end{equation}}
\newc{\barr}{\begin{eqnarray}}
\newc{\earr}{\end{eqnarray}}

	\def\sbf{\mbox{\boldmath $\sigma$}}

\def\sigmabf{\mbox{\boldmath $\sigma$}}

\begin{document} 
\topmargin -0.50in
\title {Estimating the flux of the 14.4 keV solar axions} 
\author{ F.T.  Avignone III$^{1}$, R. J. Creswick $^1$, J. D. Vergados$^{2}$, P. Pirinen$^3$, P. C. Srivastava$^{4}$ and J. Suhonen$^{3}$}
\affiliation{$^{(1)}$\it University of South Carolina, Columbia, SC 29208, USA, }
\affiliation{
  $^2$CAPP, IBS ,  Yuesong, Daejeon 34144, Republic of Korea\footnote{Permanent address, University of Ioannina, Ioannina, Gr 451 10, Greece}}
\affiliation{$^3$University of Jyvaskyla, Department of Physics, P.O. Box 35, FI-40014 University of Jyvaskyla, Finland,}
\affiliation{$^4$Department of Physics, Indian Institute of Technology, Roorkee 247667, India}
\begin{abstract}
In this paper we present a calculation of the expected flux of 
 the mono-energetic 14.4 keV solar axions emitted by the M1 type  nuclear transition of $^{57}$Fe in
the Sun. These axions  can be  detected, e.g., by inverse coherent Bragg-Primakoff conversion in single-crystal TeO$_2$
bolometers. The ingredients of this calculation are i) the axion nucleon coupling, estimated in several  popular axion models and ii)the nuclear spin matrix elements involving realistic shell model calculations with both proton  and neutron excitations. For the benefit of the experiments we have also calculated the branching ratio involving axion and photon emission. We find the solar axion flux on Earth to be $ \Phi_{a}=0.703\times 10^{9}
\mbox{cm}^{-2}\mbox{s}^{-1}\left (\frac{10^7\mbox{GeV}}{f_a}\right )^2$ and the branching ratio of axion to photon  for the same model to be:
$\frac{\omega_a}{\omega_{\gamma}}=0.229 \times 10^{-15}\approx 2\times 10^{-16}$
\end{abstract}

\date{\today}

\maketitle
\section{Introduction}
The strong CP problem in Quantum Chromodynamics (QCD), predicts the electric dipole
moment of the neutron to be much larger than the observed upper limit \cite{edipmomlim06,edipmomlim14}. Peccei and
Quinn \cite{PecQuin77b} devised an elegent solution by introducing a new U(1)PQ global symmetry that
is spontaneously broken at an energy scale $f_a$. A consequence of this U(1)PQ symmetry breaking
is that a new neutral spin-zero pseudoscalar particle (Nambu-Goldstone boson),
the axion, is generated \cite{SWeinberg78,Wilczek78}.  The axion acquires a mass through non-perturbative QCD
effects, which is inversely proportional to the symmetry-breaking parameter $f_a$. The “standard axion” with $f_a$ at the electroweak (EW) scale of 250 GeV had a convenient mass, which made it easily detectable, but was quickly excluded by early searches. This led to models with much higher value of $f_a$, which made the axions long-lived and very weakly coupled to photons, nucleons, electrons and
quarks, which makes them difficult to detect directly. This leads to the notion of "`invisible axions"'. The two most widely cited models of invisible axions are the KSVZ (Kim, Shifman, Vainshtein and Zakharov) or hadronic axion models \cite{KSVZKim79,KSVZShif80}
and the  DFSZ (Dine, Fischler, Srednicki and Zhitnitskij) or GUT axion model \cite{DineFisc81, DFSZhit80}.  This also led  to the interesting scenario of the  axion being a candidate for dark matter in the universe \cite{AbSik83,DineFisc83,PWW83,Davis86} and it can be searched for by real
experiments \cite{ExpSetUp11b,Duffy05,ADMX10,IrasGarcia12}. The relevant phenomenology has recently been reviewed \cite{PhysRep16}.
Since axions, or more generally, axion-like particles (ALPs), can couple with electromagnetic
fields or directly with leptons or quarks, the Sun could be an excellent axion source.
Solar axions could be generated by Primakoff conversion of photons, by Bremsstrahlung processes,
by Compton scattering, by electron atomic recombination, by atomic de-excitation, and by
nuclear M1 transitions. Axions produced in nuclear processes are mono-energetic because their
energies correspond to the energy difference of a specific nuclear transition. These axions can
be emitted and escape from the solar core due to the very weak interaction between the axion
and matter. Searches for solar axions have been carried out with various experimental techniques: magnetic helioscopes \cite{CAST09,CAST14},
low temperature bolometers \cite{Bolom13} and thin foil nuclear targets \cite{ThinFoil07}. CUORE(Cryogenic Underground
Observatory for Rare Events) \cite{CUORE04,CUORE05,CUORE17,Alduino17} is designed to search for neutrinoless double beta
decay ($0\nu\beta \beta$) using a very low background low temperature bolometric detector. CUORE can
also be used to search for dark matter weakly interacting massive particles (WIMPs) and solar axions \cite{LiAvigWang16}. 

In this paper we calculate
the expected  rate of 14.4 keV solar axions produced in the M1 nuclear transition
of $^{57}$Fe. We also calculate the axion to photon branching ratio and the axion  flux on Earth. To this end  in section II we discuss a set of parameters  $g_{aN}$ given  in various axion models and in section III  we present a realistic shell model calculation containing both proton and neutron excitations. In section IV  we combine these  results to get the effective matrix element to be used in section V to yield  the rate for axion production. In section VI we discuss the axion to photon branching ratio and in section VII we calculate the axion flux on Earth, using appropriate density and temperature profiles for the Sun. From 
 this flux, the axions can  be detected via the coherent inverse Primakoff process in TeO$_2$ single crystals. A brief summary of our results is given in section VIII.
\section{The particle model}
\label{sec:pmodel}
The axion $a$ is a pseudoscalar particle. Its coupling to the quarks is given by:
\beq
{\cal L}=  \frac{g_q}{f_a}i\partial_{\mu}a\bar{\psi}({\bf p}',s)\gamma^{\mu}\gamma_5\psi({\bf p}, s),
\eeq
where $g_q$ is a coupling constant and $f_a$ a scale parameter with the dimension of energy. The space component, $\mu\ne 0$, in the non relativistic limit is given by
\beq
{\cal L}=\langle \phi|\Omega|\phi \rangle,\,\Omega=\frac{g_{aq}}{2 f_a}\sbf.{\bf k},
\eeq
with $\sbf$ the Pauli matrices and  $\phi$  the quark wave function and ${\bf k}$ is the axion momentum.

We will concentrate on the last term involving the operator  $g_{aq}\sigma$. The quantities $g_{aq}$ can be evaluated at various axion models.
 \beq
\mbox{me}_q=\langle q|\left ( g_{aq}^0+g_{as}+g_{aq}^3\tau_{3}\right ) \sigma |q\rangle, 
\eeq
where we have ignored the contribution of heavier quarks and
\beq
 g_{aq}^3=\frac{1}{2}(g_{au}-g_{ad}),\,g_{aq}^0=\frac{1}{2}(g_{au}+g_{ad}).
\eeq
Some authors use the notation $c_q$ instead of $g_{aq}$.\\
Then, following a procedure analogous for the determination of the nucleon spin from that of the quarks \cite{QCDSF12,LiThomas15},  the  matrix element at the nucleon  level can  be written as:
\beq
\mbox{me}_N=\langle N |\left ( g_{aq}^0(\delta_0-\Delta s)+g_{as}\Delta s+g_{aq}^3\tau_{3}\delta_1\right )\sigma | N\rangle,
\eeq
 where
\begin{align} &\delta_0=\hspace{2pt}(\Delta u+\Delta
d+\Delta s), \nonumber \\
& \delta_1=\hspace{2pt}(\Delta u-\Delta
d),
\label{Eq:a0a1L}
\end{align} 
$\delta_0-\Delta s=\Delta u+\Delta
d$. The quantities $ \Delta u,\, \Delta d$ and $ \Delta s$ will be given below.\\
 Alternatively these quantities can be expressed in terms of the quantities $D$ and $F$  defined by Ellis \cite{JELLIS} 
$$\delta_1=F+D,\,\delta_0=3F-D+3\Delta s.$$ 
The quantity $\delta_1$ is essentially fixed by the axial current to be approximately 1.24. No such constraint exists for the isoscalar part, so for that we have to rely on models. For the quantities $D$ and $F$ one can use experimental information\cite{JELLIS}. Thus, e.g., from hyperon beta decays and flavor SU(3) symmetry one gets
$$\frac{3F - D}{\sqrt{3}}= 0.34 \pm 0.02.$$ On the other hand measurements of
$\nu p$  and $\nu \bar{p}$ elastic scattering   of the recent MicroBooNE experiment \cite{MicroBooNE17} indicate that $\Delta s = \pm 0.036\pm 0.003.$\\
The following model parameters are going to be considered:
\begin{itemize}
\item[i)] The quantities $\Delta q$ recently obtained in \cite{CHVV16}, which are consistent with lattice gauge calculations \cite{Alexandrou15}  
\beq 
\Delta u=0.897(27),\, \Delta d=-0.376(27),\, \Delta s=-0.026(4),
 \eeq
which yield
\beq
\delta_0=0.495,\, \delta_1=1.273,\,\Delta s=-0.026\Rightarrow D=0.812,\, F=0.462.
\label{Eq:a0a1L1}
\eeq
These are also  consistent with the recent results \cite{Alexandrou16}  for both connected and disconnected contributions, which yield
$$ \Delta u=0.826, \quad \Delta d=-0.386, \quad \Delta s=-0.042.$$
From these we obtain:
\beq
\delta_0=0.398,\quad \delta_1=1.212,\,\Delta s=-0.042\Rightarrow D=0.778,\quad F=0.434.
\label{Eq:a0a1L1b}
\eeq
\item [ii)] The quantities $\Delta q$ prescribed by Ellis \cite{JELLIS}, namely\\
$$\Delta u=0.78 \pm 0.02, \quad \Delta d=-0.48\pm 0.02,\quad  \Delta s=-0.15.
\pm 0.02,$$
From these we find \cite{Te125.16}
\beq
\delta_0=0.15,\quad \delta_1=1.26,\,\Delta s=-0.15 \Rightarrow D=0.795,\quad F=0.461.
\label{Eq:a0a1A}
\eeq
The small isoscalar part is consistent with the so-called proton spin crisis, i.e. the observation that the spin of the nucleon comes mainly from the gluon spins not the quark spin (EMC effect)\cite {EMC83,EMC90,EMC95} . The isovector, as we have already mentioned,  is well known  from weak interaction theory.
\item [iii)] The quantities $\Delta q$   of a recent analysis \cite{QCDSF12,LiThomas15},  found also appropriate for pseudoscalar couplings to quarks \cite{ChengChiang12}, namely 
$$\Delta u=0.84 , \Delta d=-0.43 \Delta s=-0.02 \Leftrightarrow D=0.86,\,F=0.41.$$ Thus 
\beq
\delta_0=0.43,\,\delta_1=1.27,\,\Delta s=-0.02.
\label{Eq:a0a1B}
\eeq
\end{itemize}
Thus the effective nucleon coupling becomes:
\beq
C_p=g_{aq}^0(\delta_0-\Delta s)+g_{as}\Delta s+g_{aq}^3\delta_1,\,C_n=g_{aq}^0(\delta_0-\Delta s)+g_{as}\Delta s-g_{aq}^3\delta_1
\eeq
Since $\Delta s $ is small, it seems that the largest uncertainty comes from the determination of the parameters $g_{aq}^0$ and $g_{aq}^3$. If these happen to be comparable, the smallness of $\delta_0$ makes the isoscalar contribution negligible, i.e.
\beq
C_p^{\mbox{\tiny{eff}}}=-C_n^{\mbox{\tiny{eff}}}=g_{aq}^3\delta_1.
\eeq
As a result one has in this special case essentially one unknown parameter. This, however, is not realized in the axion models considered below. As a result a relevant effective quark couplings cannot be extracted from experiment in the presence of both proton and neutron components in the nuclear wave functions.
\section{Shell-model calculation}

In the present work we have performed a shell model calculation for $^{57}$Fe in $fp$ model space using a truncation with the 
KB3G \cite{A.Poves,kbthreea,kbthreeb} effective interaction. For protons and neutrons we put no restriction for the $f_{7/2}$, $p_{3/2}$,  and $p_{1/2}$ orbitals, but we only allow a maximum of two particles in the $f_{5/2}$ orbital.
The shell model code NuShellX@MSU \cite{nushellx} was used for diagonalization of matrices.
The interaction KB3G  \cite{A.Poves} is a monopole-corrected version of the previous KB3 \cite{kbthreea,kbthreeb} interaction  in order to treat properly
the $N = 28$ and $Z = 28$ shell closures and their surroundings.
The single-particle energies for the KB3G effective interaction are taken to be -8.6000, -6.6000,
-4.6000 and -2.1000 MeV for the $f_{7/2}$, $p_{3/2}$, $p_{1/2}$ and $f_{5/2}$ orbits, respectively. 

\begin{figure}[htbp]
\center
\includegraphics[width=0.5\textwidth]{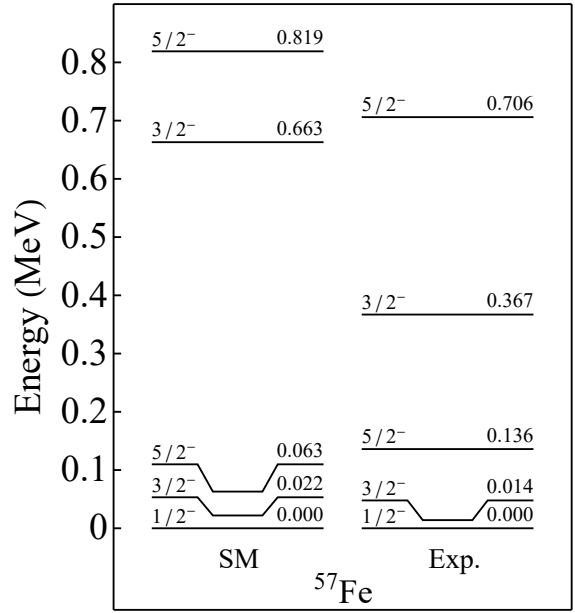}
\caption{Computed and experimental energy spectra for $^{57}$Fe up to 1 MeV.} 
\label{fig:espectra}
\end{figure}

The computed and experimental energy spectra are given in Fig. \ref{fig:espectra}. The present calculation correctly produces the $1/2^-$ as a ground state and the $3/2^-$ as the first excited state, with the predicted excited state energy  at 22 keV. The
difference, however, between the experimental result of $14.4$ keV and shell model result is only 7.6 keV. Also the order of the five lowest energy states is reproduced by our calculation, although there is a considerably wider gap between the $5/2^-_1$ and $3/2^{-}_2$ states in the shell-model spectrum than in the experimental one.
In Table \ref{tab:moments}, we  report the comparison of experimental and shell model results of 
 magnetic  and quadrupole moments. In these calculations we have used bare $g$-factors and effective charges $e_p=1.5e$ and $e_n=0.5e $. The obtained results show good agreement with available experimental data. Furthermore, we have computed the transition strength $B(\mathrm{M1}:\: 3/2^-_1 \rightarrow 1/2^-_\mathrm{gs})$ to be 0.0356 W.u., while the experimental value is 0.0078(3) W.u. For the E2 transition we computed $B(\mathrm{E2}:\: 3/2^-_1 \rightarrow 1/2^-_\mathrm{gs}) = 0.487$ W.u., while the experimental value is 0.37(7) W.u.. We find transition probabilities of
$T(\mathrm {M1}:\: 3/2^-_1 \rightarrow 1/2^-_\mathrm{gs}) = 3.4 \: \mathrm{\upmu s ^{-1}} \:$, and $T(\mathrm{E2}:\: 3/2^-_1 \rightarrow 1/2^-_\mathrm{gs}) = 4.8 \: \mathrm{ s ^{-1}} \:$, for the M1 and E2 transitions respectively. We note that the lowest transition is dominantly M1. Based on the above comparisons with experimental data, we are confident that the magnetic properties and transitions between the lowest lying states are described within reasonable accuracy by the present calculations.

\begin{table}[htbp]
\begin{center}
    \caption{Comparison of experimental \cite{ENSDF} and calculated shell model magnetic dipole moments in units of $(\mu_N)$, using $g_s^{\mathrm{eff}}=g_s^{\mathrm{free}}$ (columns 3 and 4), and electric quadrupole moments in units of ($eb$), using $e_p=1.5e$ and $e_n=0.5e$ (columns 5 and 6).} 
    \label{tab:moments}
 \begin{tabular}{  l l l l l l  }
  \hline\hline

  State \hspace{8mm}   & $E_x$ (keV)  \hspace{8mm}    &  $\mu_{\rm expt}$ \hspace{8mm}            & $\mu_{\rm SM}$  \hspace{8mm}      & $Q_{\rm expt}$ \hspace{8mm}    & $Q_{\rm SM}$        \\
\hline
 $1/2^-$     & 0   &  $+0.09044(7)$    & +0.130  & - & - \\
 $3/2^-$     & 22  &  $-0.1549(2)$     & $-0.388$  & $+0.082(8)$   &  $+0.17$  \\
\hline
\hline

\end{tabular}
 \end{center}
\end{table}

\section{The nuclear matrix element (ME)}
With the above information one can proceed to calculate the rate for axion production in the spin induced de-excitation of a nuclear level $J_i$ to a final level $J_f$. Then, the nuclear matrix element, obtained by summing over the final  and averaging over all initial m-substates, becomes
\barr
{\cal M}^2&=&\frac{1}{2J_i+1}\sum_{M_f,M_i}\left |\langle J_f M_f|\sum_n\left ( g_{aq}^0(\delta_0-\Delta s)+g_{as}\Delta s+g_{aq}^3\tau_{3}(n)\delta_1\right )\sigmabf(n).{\bf k}|J_i M_i\rangle \right |^2\nonumber\\&=&\frac{1}{2J_i+1}\frac{1}{3}|\mbox{ME}|^2,
\label{Eq:FullME}
\earr
where
\beq
\mbox{ME}=\langle J_f ||\sum_n\left( g_{aq}^0(\delta_0-\Delta s)+g_{as}\Delta s+g_{aq}^3\tau_{3}(n)\delta_1\right )\sigmabf(n)||J_i\rangle .
\eeq
The last expression is the usual reduced matrix of the one body operator in the isospin basis. The sum over $n$ involves all active nucleons.\\
Alternatively, the reduced matrix element can be computed in the proton-neutron representation
\beq
\Omega_p=\langle J_f ||\sum_n\left ( \frac{1}{2}(1+\tau_{3}(n))\right )\sigmabf(n)||J_i \rangle,\,\Omega_n=\langle J_f ||\sum_n\left ( \frac{1}{2}(1-\tau_{3}(n))\right )\sigmabf(n)||J_i\rangle.
\eeq
Thus the reduced ME can be written as:
\beq
\mbox{ME}=C_p \Omega_p+C_n \Omega_n.
\eeq
Our shell model calculation for the $^{57}$Fe  transition predicts:
\beq
\Omega_p=0.1054,\quad\Omega_n=0.7932,\,
\eeq
As expected the neutron component dominates, but the proton may make an important contribution for some coupling choices.\\ 
Our shell model calculation also predicts the matrix element of the  isovector orbital magnetic moment operator, namely:
\beq
\langle J_f||\sum_k \ell(k) \tau_3(k)||J_i\rangle=0.8291,
\eeq
which enters the de-excitation of the 14.4 keV state via photon emission, which is needed to estimate the axion to photon branching ratio  discussed below. \\
  To proceed further in the evaluation of the  MEs, we need some elementary particle input, e.g. those  recently obtained for the hadronic model of the recent calculation\cite{CHVV16}:
$$
g_{au}=-0.47,g_{ad}=0.02\Rightarrow g_{aq}^0=-0.2250,g_{aq}^3=-0.2465.
$$ 
Combining this with the $\delta_0$, $\Delta_s$ and $\delta_1$ given by Eq.(\ref{Eq:a0a1L1})
we get
$$
 C_p=(g_{aq}^0(\delta_0-\Delta_s)+g_{aq}^3\delta_1)=-0.429,\,C_n=(g_{aq}^0\delta_0-\Delta_s)-g_{aq}^3\delta_1)=0.195 \mbox{ (MODEL A)}.
$$
On the other hand using Eqs (\ref{Eq:a0a1A}) and (\ref{Eq:a0a1B}) we get
$$ C_p=-0.378, C_n=0.241 \mbox{ (MODEL B)};\, C_p=-0.404, C_n=0.219 \mbox{ (MODEL C)}.$$
Finally if one considers renormalization effects in the KSVZ model these authors conclude \cite{CHVV16}
$$C_p^{\mbox{{\tiny KSVZ}}}=-0.47,\,C_n^{\mbox{{\tiny KSVZ}}}=-0.02 \mbox{ (MODEL D)}.$$
Models $A, B, C$ lead to large values  for the neutron coupling, but in the presence of renormalization 
 the neutron contribution is greatly suppressed and even its sign changes. So the first three models are included for orientation purposes, but they should not be taken seriously.
 
 The coupling of axion to matter has been  investigated \cite{CHVV16, RingSaika15}, in particular in  the context of the DFSZ axion models \cite{DineFisc83, DFSZhit80}. 
At the quark level they find :
$$g_{au}=g_{ac}=g_{at}=\frac{1}{3}\sin^2{\beta},\,g_{ad}=g_{as}=g_{ab}=\frac{1}{3}-\frac{1}{3}\sin^2{\beta}$$
Thus restricting ourselves to quarks $u$, $d$ and $s$, which are relevant for nucleons, we get:
$$ g^0_{aq}=\frac{1}{6} ,\, g_{as}=\frac{1}{3}-\frac{1}{3}\sin^2{\beta}, \, g^1_{aq}=-\frac{1}{6}+\frac{1}{3}\sin^2(\beta),$$
where $\tan{\beta}$ is the ratio of the vacuum expectation values of the two doublets of the model, which, like the well known case  in supersymmetry, is not constrained by the SM physics.\\
 In this case we will explore the dependence of the results on the quark model and  $\tan{\beta}$. In a fashion analogous to  the constrained parameter space of supersymmetry, we will consider values of $\tan{\beta}$, e.g. $\tan{\beta}=10$  as well as  a  small value, e.g. $\tan{\beta}=1$. 
 The obtained results are presented in table \ref{tab:DSFZ}.
\begin{table}
\begin{center}
\caption{The results obtained in the DSFZ models for  various isoscalar and isovector contributions and two choices of $\tan{\beta}$ }
\label{tab:DSFZ}
$$\begin{array}{c|cccc|cccc}
& & &\tan{\beta}=1&&&\tan{\beta}=10&\\
&Eq. (\ref{Eq:a0a1L1})&Eq. (\ref{Eq:a0a1L1b})&Eq. (\ref{Eq:a0a1A})&Eq. (\ref{Eq:a0a1B})&Eq. (\ref{Eq:a0a1L1})&Eq. (\ref{Eq:a0a1L1b})&Eq. (\ref{Eq:a0a1A})&Eq. (\ref{Eq:a0a1B})\\
\hline
  C_p&0.0825 & 0.0663 & 0.025 & 0.065 & 0.2947 &
   0.2712 & 0.2553 & 0.2757 \\
 C_n& 0.0825 & 0.0663 & 0.025 & 0.065 & -0.1212 &
   -0.1248 & -0.1563 & -0.1392 \\
  \mbox{ME}&0.074 & 0.0596 & 0.0225 & 0.0584 & -0.0651 &
   -0.0704 & -0.0971 & -0.0814 \\
 \frac{{\cal M}^2}{10^{-3}}&0.458 & 0.2961 & 0.0421 & 0.2843 & 0.353 &
   0.4131 & 0.7856 & 0.5516 \\
\end{array}
 $$
\end{center}
\end{table}

From table  \ref {tab:DSFZ} we make a reasonable selection  for our calculations 
\barr
C_p&=&0.0663,C_n=0.0663\mbox{ for }\tan{\beta}=1 \mbox{ MODEL E},\nonumber\\C_p&=&0.2712,C_n=-0.1248 \mbox{ for }\tan{\beta}=10 \mbox{ MODEL F}.
\earr
Combining these with the reduced nuclear matrix elements for protons and neutrons we obtain:
$$\mbox{ME}=0.0596 \mbox{ for }\tan{\beta}=1 \mbox{ MODEL E},\,\mbox{ME}=-0.0704\mbox{ for }\tan{\beta}=10 \mbox{ MODEL F} $$
With supersymmetry as a guide we expect the larger value of $\beta$ as the most likely.   

Before ending this exposition we should mention that there exist  a fairly old model which depends on essentially only one family parameter developed long time ago by Kaplan \cite{Kaplan85}, based on the DFSZ \cite{DineFisc83, DFSZhit80} axion. In this case there exists  one value $c_q$ indicated by $(1/2)X_u$ for the charge 2/3 quarks and one  indicated by $(1/2)X_d$ for the charge -1/3 quarks, with the condition $X_u+X_d=1$, $X_u>0$ and $X_d>0$. 
At the nucleon level the isovector contribution depends on $c_q$, but the  isoscalar contribution becomes independent of $c_q$, in other words:
$$g_{aq}^0=\frac{1}{4},\, g_{as}=\frac{1}{2}x_d,\,g_{aq}^3=\frac{1}{4}(2 X_u-1),g_{aq}^3\le g_{aq}^0.$$
We will consider the isoscalar and isovector component found in cases i) ii) and iii) above (see section \ref{sec:pmodel}) and the following 3 quark couplings $X_u$ and $X_d$:
\begin{itemize}
\item[a)] $X_u=X_d=1/2$
Then we obtain:
$$ \begin{array}{c|cccc}&Eq. (\ref{Eq:a0a1L1})&Eq. (\ref{Eq:a0a1L1b})&Eq. (\ref{Eq:a0a1A})&Eq. (\ref{Eq:a0a1B})\\
\hline\\
Cp& 0.2540 & 0.2055 & 0.1125&0.2000\\
   \\
C_n& 0.2540 & 0.2095 & 0.1125&0.2000\\
   \\
	\end{array}
	$$
	\item [b)] $X_u=3/4,X_d=1/4$ Then
	$$
	\begin{array}{r|cccc}&Eq. (\ref{Eq:a0a1L1})&Eq. (\ref{Eq:a0a1L1b})&Eq. (\ref{Eq:a0a1A})&Eq. (\ref{Eq:a0a1B})\\
	\hline
 C_p&0.4165 & 0.3663 & 0.2888&0.3613\\
C_n& 0.0981 & 0.0633& 0.0213&0.0438 \\
\end{array}
$$
\item [c)] $X_u=1/4,X_d=3/4$ Then
$$
\begin{array}{r|ccrc}&Eq. (\ref{Eq:a0a1L1})&Eq. (\ref{Eq:a0a1L1b})&Eq. (\ref{Eq:a0a1A})&Eq. (\ref{Eq:a0a1B})\\
\hline
 C_p&0.09163 & 0.0523&-0.0638 &0.0388 \\
 C_n&0.4099 & 0.3557 & 0.2513&0.3565\\
   \\
\end{array}.
$$
\end{itemize}
From the above set we select  three typical cases:
\barr
C_p&=&C_n=0.1125 \mbox{ (MODEL G) };C_p=0.3663,C_n=0.0633 \mbox{ (MODEL H) };\nonumber\\C_p&=&0.0388,C_n=0.3565 \mbox{ (MODEL I) }.
\label{Eq:XuXd}
\earr
Since the suppression of the neutron coupling looms on the horizon,  we had to consider an elaborate shell model calculation for $^{57}$Fe in which the proton components of the wave function may be involved and contribute significantly to the width for axion production. The obtained results are given in table \ref{tab:tabme}.
\begin{table}
\begin{center}
\caption{The various couplings and matrix elements entering axion production
\label{tab:tabme}}
$$
\begin{array}{c|rrrcc}MODEL&C_p&C_n&\mbox{ME}&{\cal M}^2&|\mbox{rME}|^2\\
\hline
A& -0.4291 & 0.1947 & 0.1092 & 0.000993 &
   0.1008 \\
B& -0.3762 & 0.2412 & 0.1517 & 0.001917&
   0.1945 \\
 C&-0.4034 & 0.2189 & 0.1311 & 0.001433 &
   0.1454 \\
D& -0.4700 & -0.0200 & -0.0654 & 0.000356 &
   0.0362 \\
 E&0.0663 & 0.0663& 0.0596 & 0.000296
   & 0.0300 \\
 F&0.2712 & -0.1248 & -0.0704 & 0.000413
   & 0.0419 \\
	G& 0.1125 & 0.1125 & 0.1011 & 0.000852 &
   0.0864 \\
H& 0.3663 & 0.0633 & 0.0888 & 0.000657 &
   0.0667 \\
I& 0.0388 & 0.3563 & 0.2867 & 0.006848 &
   0.6948 \\
\end{array}
$$
\end{center}
\end{table}
The quantity  $|\mbox{rME}|^2 $ enters in the axion to photon production branching ratio to be discussed below.
\section{The rate of axion production}
 Once the Matrix element ${\cal M}^2$, see Eq. (\ref{Eq:FullME}),  is known  the axion production width is given by
\beq
\Gamma(J_i\rightarrow J_f a)=\frac{1}{2\pi}\frac{\left (\sqrt{\Delta^2-m_a^2}\right )^3}{4 f_a^2}{\cal M}^2,
\eeq
where as we have seen in the previous section the expression for ${\cal M}^2 $ involves the needed  particle and nuclear physics inputs.\\
The mass of the axion is expected to be much less than the transition energy  $\Delta$ \cite{CHVV16}, e.g. $m_a=5.7\mu\mbox{eV}\times\frac{1}{(f_a/10^{12}\mbox{GeV})}.$ 
Even for the unrealistically small scale of $f_a=10^{6}$ GeV we find $m_a= 5.7\mbox{eV}$. Thus
\barr
\Gamma(J_i\rightarrow J_f a)&=&\Lambda \left (\frac{\Delta}{\mbox{14.4 keV}}\right )^3,\nonumber\\ 
&& \Lambda=1.2\times 10^{-21}\mbox{eV}  \left (\frac{10^{7}\mbox{GeV}}{f_a}\right )^2{\cal M}^2=1.2 \times 10^{-7}\mbox{eV}  \left  (\frac{f_a}{1\mbox{GeV}}\right )^{-2}{\cal M}^2
\earr
or
\barr
\Lambda&=&1.83 \times 10^{-6}\mbox{s}^{-1}  \left (\frac{10^{7}\mbox{GeV}}{f_a}\right )^2{\cal M}^2
=1.83 \times 10^{8}\mbox{s}^{-1}  \left  (\frac{f_a}{1\mbox{GeV}}\right )^{-2}{\cal M}^2.\nonumber\\
\earr
\\ 
Using the ${\cal M}^2$ obtained above and the excitation energy of $\Delta=14.4$ keV we find the width and transition probability as follows:
$$
\begin{array}{c|ccccccccc}
\mbox{MODEL}&A&B&C&D&E&F&G&H&I\\
\hline\\
\frac{\Gamma}{(10^{-10})}\mbox{eV}\left  (\frac{f_a}{1\mbox{\tiny{GeV}}}\right )^{-2}&1.19&2.30&1.72&0.4282&0.355&0.496&1.022&
   0.788&8.22\\
	\frac{\omega_a}{(10^{5})}\mbox{s}^{-1}\left  (\frac{f_a}{1\mbox{\tiny{GeV}}}\right )^{-2}&1.82&3.51&2.62&0.652&0.542&0.756&1.56&1.20&12.5\\
\end{array}
$$
\section{The axion to photon production branching ratio}
Experimentally it is of interest to estimate the axion to photon branching ratio. Following Haxton and Lee \cite{HaxLee91} and  \cite{Peccei96} we write
\barr
\frac{\omega_a}{\omega_{\gamma}}&=&\frac{1}{2 \pi \alpha}\left (\frac{m_N}{2f_a}\right )^2
\frac{ |\mbox{rME}|^2}{\left ((\mu_0-\frac{1}{2}\beta+\mu_1-\eta \right )^2}
,\nonumber\\ \mbox{rME}&=&\beta(C_p+C_n)+C_p-C_n,
\earr
where $ m_N$ is the nucleon mass, $\mu_0=0.88$, $\mu_1=4.77$, $$\beta=\frac{(\Omega_p+\Omega_n)}{(\Omega_p-\Omega_n)},\quad \eta=-\langle J_f||\sum_k \ell(k) \tau_3(k)||J_i\rangle/\langle J_f||\sum_k \sigma(k) \tau_3(k)||J_i\rangle.$$
The quadrupole transition contribution $\delta$ is negligible in this case, but it may have to be included in other experimentally interesting nuclei, like $^{65}$Cu \cite{Avignone88}. 
Our nuclear calculation yields  $\beta=-1.3065$ and $\eta=1.2054$. i.e. $\left (\mu_0-\frac{1}{2}\right )\beta+\mu_1-\eta=3.068$. 
This must be compared with the value 3.458 of  Haxton and Lee \cite{HaxLee91}. We thus get 
\beq
 \frac{\omega_a}{\omega_{\gamma}}=0.58 \left (\frac{M_N}{f_a}\right )^2  \left (rME\right )^2.
\label{Eq:romega1}
\eeq
This equation can be used to extract a limit on the parameter $\left (rME\right )^2$ from the branching ratio data.
 The values of  $|\mbox{rME}^2|$ for the models considered in this work are given in table \ref{tab:tabme}. Using these parameters and for the value   $f_a= 10^7\mbox{GeV}$, see section \ref{sec:AxionFlux}, we obtain:
\beq
\begin{array}{c|ccccccccccc}
&&&&&\mbox{MODEL}&&&&&&\\
\hline
&A&B&C&D&E&F&G&H&I\\
\hline
\left(\frac{\omega_a}{\omega_{\gamma}}\right)/\left ( 10^{-15}\right )&0.548& 1.060& 0.794& 0.197& 0.164& 0.229& 0.472& 0.363& 3.786\\
\end{array}
\label{Eq:romega2}
\eeq

\section{Axion Flux in the Earth}
\label{sec:AxionFlux}
\begin{figure}
  \begin{center}
\rotatebox{90}{\hspace{-0.0cm} $ \rho\left (\frac{r}{R_{\odot}}\right )\rightarrow$}
\includegraphics[width=0.8\textwidth,height=0.3\textwidth]{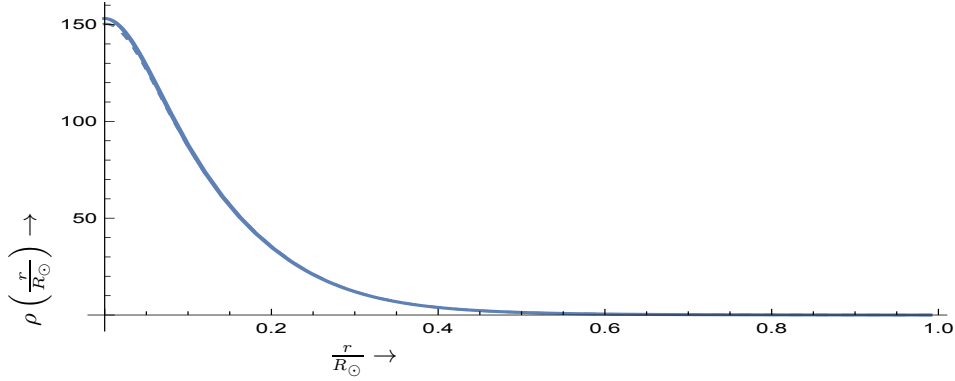}\\
{\hspace{-3.0cm}$ \frac{r}{R_{\odot}}\rightarrow$  }
 \caption{The solar density profiles in two models described in Ref. \cite{Bahcall06}.
 \label{fig:SolarDen}}
 \end{center}
  \end{figure}

\begin{figure}
  \begin{center}
\rotatebox{90}{\hspace{-0.0cm} $ T\rightarrow ^{0}\,K$}
\includegraphics[width=0.8\textwidth,height=0.3\textwidth]{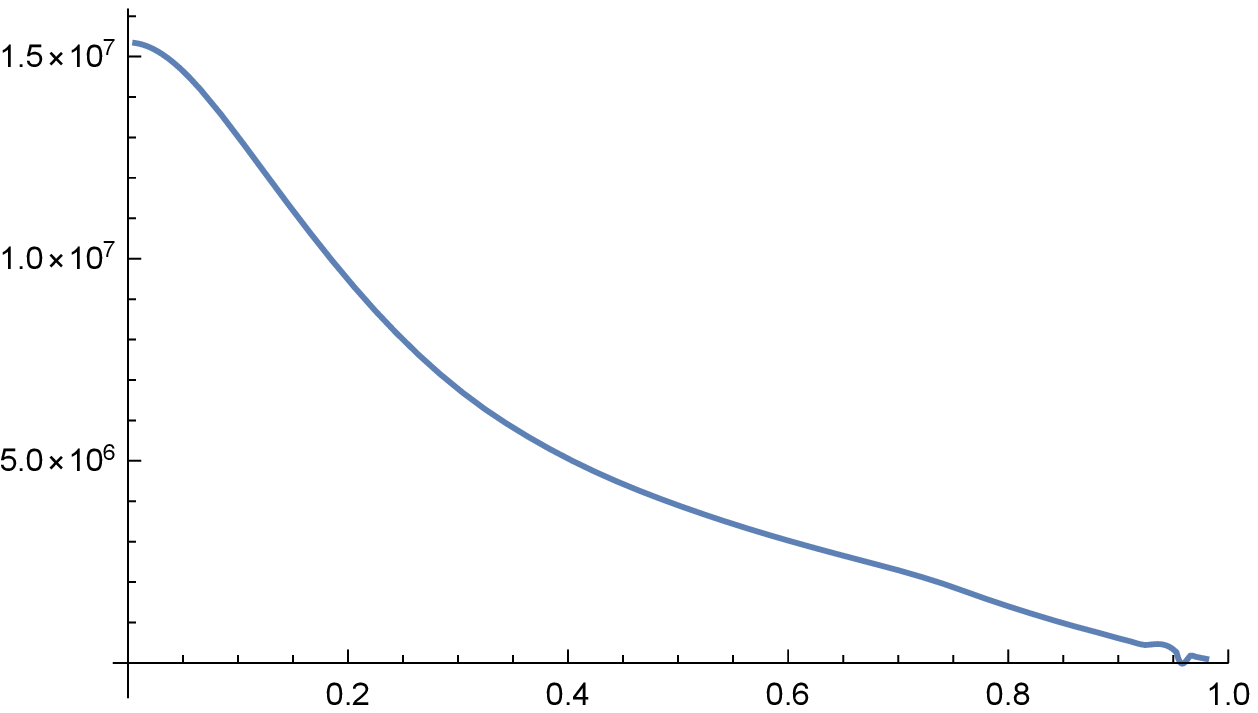}\\
{\hspace{-3.0cm}$ \frac{r}{R_{\odot}}\rightarrow$  }
 \caption{The solar temperature profile obtained from the data of Ref. \cite{SolarT92}.
 \label{fig:SolarT}}
 \end{center}
  \end{figure}

One must first estimate the number of $^{57}$Fe in the excited state with $J_1=3/2$. This is given by the Boltzmann factor
$$\frac{N^{*}}{N}=\frac{2 J_1+1}{2 J_0+1}e^{-\frac{\Delta}{kT}},$$
where $J_0=1/2$ is the ground state  angular momentum,  $N$ is the number of $^{57}$Fe present  and $T$ the temperature of the Sun. This fraction depends on the position in the Sun. Thus we can write \cite{SFluxPR93,SFluxCAST09}
\beq
\frac{N^{*}}{N}=\frac{2 J_1+1}{2 J_0+1}fr=2 fr,\,fr=\frac{\int_0^{R_{\odot}} e^{-\frac{\Delta}{kT(r)}} 4 \pi r^2 \rho(r)dr}{\int_0^{R_{\odot}}4 \pi r^2 \rho(r)dr},
\eeq 
where $\rho(r)$ and $T(r)$ are the density and temperature profiles of the Sun respectively. We found it convenient to express the above expression in units of $x=\frac{r}{R_{\odot}}$ and write:
\beq
fr=\frac{\int_0^{1} e^{-\frac{\Delta}{kT(x)}} 4 \pi x^2 \rho(x)dx}{\int_0^{1}4 \pi x^2 \rho(x)dx}.
\eeq
To proceed further we need the temperature and the density profile of the Sun. The density profile \cite{Bahcall06} is presented in Fig. \ref{fig:SolarDen}.
A solar temperature profile can be found in Ref. \cite{SolarT92} and is exhibited in
Fig. \ref{fig:SolarT}.
For the solar density profiles in two models described in Ref. \cite{Bahcall06} we obtain almost the same result $fr=6.75\times 10^{-7}$ and $fr=6.71\times 10^{-7}$. On the other hand using the density and temperature files found in ref. \cite{Bahcall95} we find $fr=7.75\times 10^{-7}$.
 Adopting the first value and taking as the number of $^{57}$Fe nuclei in the Sun per gram to be  $3 \times 10^{17}\mbox{g}^{-1}$ \cite{SFluxPR93,SFluxCAST09}, we find the total number  $N^{*}$ of $^{57}$Fe  nuclei decaying  in the Sun to be
$$N^{*}=2 \times 6.75  \times 10^{-7}  \times 3 \times 10^{17} \times 2.0 \times 10^{33}=0.81 \times 10^{45}.$$
Adopting the value of $1.0 \times 10^{45}$  the rate  of the emitted axions  is:
$$N_a=N^{*} \Gamma(J_i\rightarrow J_f a) $$
and the flux of axions at the Earth is:
\beq
\Phi_a=\frac{N*}{4 \pi d^2_{SE}}{\cal M}^2 \times 1.83 \times 10^{8}\mbox{s}^{-1}  \left  (\frac{f_a}{1\mbox{GeV}}\right )^{-2}=6.86\times 10^{25}\mbox{cm}^{-2}\mbox{s}^{-1}{\cal M}^2\left  (\frac{f_a}{1\mbox{GeV}}\right )^{-2},
\label{Eq:flux1}
\eeq
where $ d_{SE}$ is the average distance between the Earth and the Sun.  Thus we find:
\beq
\begin{array}{c|ccccccccc|c|cc}
&&&&&\mbox{MODEL}&&&&&&&\\
\hline
&A&B&C&D&E&F&G&H&I&\mbox{units}&f_a\mbox{ factor}\\
\hline
\Phi_a&0.681&1.32&0.983&0.245&0.203&0.283&0.584&0.451&4.70&
\times10^{23}\mbox{cm}^{-2}\mbox{s}^{-1}& \left  
(\frac{f_a}{1\mbox{GeV}}\right )^{-2}\\ \end{array}
\label{Eq:flux2}
\eeq
in the above order of the ME.\\
The numerical value depends, of course, on the  value of $f_a$. It may be useful to relate this to the usual axion-photon coupling $g_{a\gamma \gamma}$ given by:
\beq
g_{a\gamma \gamma}=\frac{\alpha C_{\gamma}}{2 \pi f_a},
\eeq
where $C_{\gamma}$ is a model dependent parameter, which in the KSVZ model \cite{KSVZKim79,KSVZShif80} takes the value
$−1.95 \pm 0.08$.
Taking now the rather optimistic value $g_{a\gamma \gamma}=0.66\times 10^{-10}\mbox{GeV}^{-1}$, extracted from axion searches, e.g. the CAST \cite{CAST07} limit $g_{a\gamma \gamma}\leq 1.16 \times 10^{-10}\mbox{GeV}^{-1}$, and  the more recent \cite{CAST17b} CAST limit (0.$66\times 10^{−10}$ GeV$^{−1}$
at $95\%$ confidence level) as well as astrophysical limits  $g_{a\gamma \gamma}\leq 10^{-10}\mbox{GeV}^{-1}$ \cite{Raffelt07}, one obtain $f_a\geq 3.4 \times 10^{7}$GeV. We will adopt the value of $10^7$ GeV in our work. Thus the obtained flux is
\beq
\begin{array}{c|ccccccccc|ccc}
&&&&&\mbox{MODEL}&&&&&&&\\
\hline
&A&B&C&D&E&F&G&H&I&\mbox{units}&\\
\hline
\Phi_a&1.691&3.262&2.437& 0.607& 0.504& 0.703& 1.449& 1.119& 11.669&\times 10^{9}
\mbox{cm}^{-2}\mbox{s}^{-1}&\\
\end{array}
\label{Eq:flux3}
\eeq
\section{Discussion}
In this paper we have performed a calculation of the expected flux of 
 the mono-energetic 14.4 keV solar axions emitted by the M1 type  nuclear transition of $^{57}$Fe in
the Sun. To this end we have included the following ingredients:
\begin{itemize}
\item We have employed the spin induced axion quark couplings obtained in various axion models.
\item We appropriately transformed these isovector and isoscalar couplings at the nucleon model.
\item We performed a realistic shell model calculation of the 14.4 keV state of $^{57}$Fe containing both protons and neutrons
\item With the previous input we were able to obtain the nuclear matrix elements for axion production as well as the branching ratio of axion to photon production. 
\item We also obtained the number of excited nuclei found in the Sun employing a Boltzmann factor obtained with appropriate density and temperature profiles in the Sun.
\item With the above ingredients we have obtained the flux of solar axions in the surface of the Earth.
\end{itemize}
We give a formula which allows the experimentalists to extract the effective matrix element ${\cal M}^2 $ from the data, Eq. (\ref{Eq:flux1}) and then compare it to the predictions of the various axion models, Eqs. (\ref{Eq:flux2}) and (\ref{Eq:flux3}). We also give analogous formulas for the branching ratio of axions to photons  involving the matrix element  $|\mbox{rME}|^2$. A value can be extrcted from the data via Eq. (\ref{Eq:romega1}) and an estimate for the expected branching ratio is given by Eq. (\ref{Eq:romega2}). Since, however, our nuclear model involves both protons and neutrons, in extracting the above values from the data, one cannot disentangle the couplings  from the nuclear matrix elements.

As a conclusion we can say that, within  the popular  DFSZ axion model  for large $\tan{\beta}$ considered here and a realistic calculation of the nuclear matrix elements employed, involving both proton and neutron configurations, a reasonably high flux of axions is expected on earth, coming from the decay 14.4 keV state of $^{57}$Fe in the Sun, namely:
\beq
\Phi_{a}=0.703\times 10^{9}
\mbox{cm}^{-2}\mbox{s}^{-1}\left (\frac{10^7\mbox{GeV}}{f_a}\right )^2.
\eeq
Furthermore the branching ratio of axion to photon for the same model is is predicted to be:
\beq\frac{\omega_a}{\omega_{\gamma}}=0.229 \times 10^{-15}\approx 2\times 10^{-16}
\eeq
Finally it is  worth mentioning that the width of the 14.4   keV state of $^{57}$Fe  involved here is of the order of few eV. while  that of the axion distribution  due to the Primakoff effect  is of  some keV. 

\begin{thebibliography}{57}
\expandafter\ifx\csname natexlab\endcsname\relax\def\natexlab#1{#1}\fi
\expandafter\ifx\csname bibnamefont\endcsname\relax
  \def\bibnamefont#1{#1}\fi
\expandafter\ifx\csname bibfnamefont\endcsname\relax
  \def\bibfnamefont#1{#1}\fi
\expandafter\ifx\csname citenamefont\endcsname\relax
  \def\citenamefont#1{#1}\fi
\expandafter\ifx\csname url\endcsname\relax
  \def\url#1{\texttt{#1}}\fi
\expandafter\ifx\csname urlprefix\endcsname\relax\def\urlprefix{URL }\fi
\providecommand{\bibinfo}[2]{#2}
\providecommand{\eprint}[2][]{\url{#2}}

\bibitem[{\citenamefont{Bakeret et~al.}(2006)}]{edipmomlim06}
\bibinfo{author}{\bibfnamefont{C.~A.} \bibnamefont{Bakeret}}
  \bibnamefont{et~al.}, \bibinfo{journal}{Phys.Rev. Lett}
  \textbf{\bibinfo{volume}{97}}, \bibinfo{pages}{131801}
  (\bibinfo{year}{2006}), \bibinfo{note}{an Improved experimental limit on the
  electric dipole moment of the neutron}.

\bibitem[{\citenamefont{Baron et~al.}(2014)}]{edipmomlim14}
\bibinfo{author}{\bibfnamefont{J.}~\bibnamefont{Baron}} \bibnamefont{et~al.},
  \bibinfo{journal}{Science} \textbf{\bibinfo{volume}{343}},
  \bibinfo{pages}{269} (\bibinfo{year}{2014}), \bibinfo{note}{dOI:
  10.1126/science.1248213}.

\bibitem[{\citenamefont{Peccei and Quinn}(1977)}]{PecQuin77b}
\bibinfo{author}{\bibfnamefont{R.~D.} \bibnamefont{Peccei}} \bibnamefont{and}
  \bibinfo{author}{\bibfnamefont{H.~R.} \bibnamefont{Quinn}},
  \bibinfo{journal}{Phys. Rev. D} \textbf{\bibinfo{volume}{16}},
  \bibinfo{pages}{1791} (\bibinfo{year}{1977}).

\bibitem[{\citenamefont{Weinberg}(1978)}]{SWeinberg78}
\bibinfo{author}{\bibfnamefont{S.}~\bibnamefont{Weinberg}},
  \bibinfo{journal}{Phys. Rev. Lett.} \textbf{\bibinfo{volume}{40}},
  \bibinfo{pages}{223} (\bibinfo{year}{1978}).

\bibitem[{\citenamefont{Wilczek}(1978)}]{Wilczek78}
\bibinfo{author}{\bibfnamefont{F.}~\bibnamefont{Wilczek}},
  \bibinfo{journal}{Phys. Rev. Lett.} \textbf{\bibinfo{volume}{40}},
  \bibinfo{pages}{279} (\bibinfo{year}{1978}).

\bibitem[{\citenamefont{Kim}(1979)}]{KSVZKim79}
\bibinfo{author}{\bibfnamefont{J.~E.} \bibnamefont{Kim}},
  \bibinfo{journal}{Phys. Rev. Lett.} \textbf{\bibinfo{volume}{43}},
  \bibinfo{pages}{137} (\bibinfo{year}{1979}).

\bibitem[{\citenamefont{Shifman et~al.}(1980)\citenamefont{Shifman, Vainshtein,
  and Zakharov}}]{KSVZShif80}
\bibinfo{author}{\bibfnamefont{M.~A.} \bibnamefont{Shifman}},
  \bibinfo{author}{\bibfnamefont{A.}~\bibnamefont{Vainshtein}},
  \bibnamefont{and} \bibinfo{author}{\bibfnamefont{V.~I.}
  \bibnamefont{Zakharov}}, \bibinfo{journal}{Nuc. Phys. B}
  \textbf{\bibinfo{volume}{166}}, \bibinfo{pages}{493} (\bibinfo{year}{1980}).

\bibitem[{\citenamefont{Dine et~al.}(1981)\citenamefont{Dine, Fischler, and
  Srednicki}}]{DineFisc81}
\bibinfo{author}{\bibfnamefont{M.}~\bibnamefont{Dine}},
  \bibinfo{author}{\bibfnamefont{W.}~\bibnamefont{Fischler}}, \bibnamefont{and}
  \bibinfo{author}{\bibfnamefont{M.}~\bibnamefont{Srednicki}},
  \bibinfo{journal}{Phys. Lett.} \textbf{\bibinfo{volume}{B 104}},
  \bibinfo{pages}{199} (\bibinfo{year}{1981}).

\bibitem[{\citenamefont{Zhitnisky}(1980)}]{DFSZhit80}
\bibinfo{author}{\bibfnamefont{A.}~\bibnamefont{Zhitnisky}},
  \bibinfo{journal}{Sov. J. Nuc. Phys.} \textbf{\bibinfo{volume}{31}},
  \bibinfo{pages}{260} (\bibinfo{year}{1980}), \bibinfo{note}{in Russian}.

\bibitem[{\citenamefont{Abbott and Sikivie}(1983)}]{AbSik83}
\bibinfo{author}{\bibfnamefont{L.~F.} \bibnamefont{Abbott}} \bibnamefont{and}
  \bibinfo{author}{\bibfnamefont{P.}~\bibnamefont{Sikivie}},
  \bibinfo{journal}{Phys. Lett.} \textbf{\bibinfo{volume}{B120}},
  \bibinfo{pages}{133} (\bibinfo{year}{1983}).

\bibitem[{\citenamefont{Dine and Fischler}(1983)}]{DineFisc83}
\bibinfo{author}{\bibfnamefont{M.}~\bibnamefont{Dine}} \bibnamefont{and}
  \bibinfo{author}{\bibfnamefont{W.}~\bibnamefont{Fischler}},
  \bibinfo{journal}{Phys. Lett.} \textbf{\bibinfo{volume}{B 120}},
  \bibinfo{pages}{137} (\bibinfo{year}{1983}).

\bibitem[{\citenamefont{Preskill et~al.}(1983)\citenamefont{Preskill, Wise, and
  Wilczek}}]{PWW83}
\bibinfo{author}{\bibfnamefont{J.}~\bibnamefont{Preskill}},
  \bibinfo{author}{\bibfnamefont{M.~B.} \bibnamefont{Wise}}, \bibnamefont{and}
  \bibinfo{author}{\bibfnamefont{F.}~\bibnamefont{Wilczek}},
  \bibinfo{journal}{Phys. Lett.} \textbf{\bibinfo{volume}{B120}},
  \bibinfo{pages}{127} (\bibinfo{year}{1983}).

\bibitem[{\citenamefont{R.L.Davis}(1986)}]{Davis86}
\bibinfo{author}{\bibnamefont{R.L.Davis}}, \bibinfo{journal}{Phys. Lett B}
  \textbf{\bibinfo{volume}{180}}, \bibinfo{pages}{225} (\bibinfo{year}{1986}).

\bibitem[{\citenamefont{Asztalos et~al.}(2010)}]{ExpSetUp11b}
\bibinfo{author}{\bibfnamefont{S.~J.} \bibnamefont{Asztalos}}
  \bibnamefont{et~al.} (\bibinfo{collaboration}{{The ADMX collaboration}}),
  \bibinfo{journal}{Phys. Rev. Lett.} \textbf{\bibinfo{volume}{104}},
  \bibinfo{pages}{041301} (\bibinfo{year}{2010}),
  \bibinfo{note}{arXiv:0910.5914 (astro-ph.CO)}.

\bibitem[{\citenamefont{Duffy et~al.}(2005)}]{Duffy05}
\bibinfo{author}{\bibfnamefont{L.}~\bibnamefont{Duffy}} \bibnamefont{et~al.}
  (\bibinfo{collaboration}{{The ADMX collaboration}}), \bibinfo{journal}{Phys.
  Rev. Lett.} \textbf{\bibinfo{volume}{95}}, \bibinfo{pages}{09134}
  (\bibinfo{year}{2005}).

\bibitem[{\citenamefont{Wagner et~al.}(2010)}]{ADMX10}
\bibinfo{author}{\bibfnamefont{A.}~\bibnamefont{Wagner}} \bibnamefont{et~al.}
  (\bibinfo{collaboration}{{The ADMX collaboration}}), \bibinfo{journal}{Phys.
  Rev. Lett.} \textbf{\bibinfo{volume}{105}}, \bibinfo{pages}{171801}
  (\bibinfo{year}{2010}), \bibinfo{note}{arXiv:1007.3766 (astro-ph.CO)}.

\bibitem[{\citenamefont{Irastorza and García}(2012)}]{IrasGarcia12}
\bibinfo{author}{\bibfnamefont{I.~G.} \bibnamefont{Irastorza}}
  \bibnamefont{and} \bibinfo{author}{\bibfnamefont{J.~A.}
  \bibnamefont{García}}, \bibinfo{journal}{JCAP}
  \textbf{\bibinfo{volume}{1210}}, \bibinfo{pages}{022} (\bibinfo{year}{2012}),
  \bibinfo{note}{arXiv:1007.3766 (astro-ph.IM)}.

\bibitem[{\citenamefont{Marsh}(2016)}]{PhysRep16}
\bibinfo{author}{\bibfnamefont{D.~J.~E.} \bibnamefont{Marsh}},
  \bibinfo{journal}{Phys. Rep. D} \textbf{\bibinfo{volume}{643}},
  \bibinfo{pages}{1} (\bibinfo{year}{2016}).

\bibitem[{\citenamefont{Andriamonje et~al.}(2009{\natexlab{a}})}]{CAST09}
\bibinfo{author}{\bibfnamefont{S.}~\bibnamefont{Andriamonje}}
  \bibnamefont{et~al.} (\bibinfo{collaboration}{{The CAST collaboration}}),
  \bibinfo{journal}{J. Cosmol. Astropart. Phys.} \textbf{\bibinfo{volume}{12}},
  \bibinfo{pages}{002} (\bibinfo{year}{2009}{\natexlab{a}}).

\bibitem[{\citenamefont{Arik et~al.}(2014)}]{CAST14}
\bibinfo{author}{\bibfnamefont{M.}~\bibnamefont{Arik}} \bibnamefont{et~al.}
  (\bibinfo{collaboration}{{The CAST collaboration}}), \bibinfo{journal}{Phys.
  Rev. Lett.} \textbf{\bibinfo{volume}{112}}, \bibinfo{pages}{091302}
  (\bibinfo{year}{2014}).

\bibitem[{\citenamefont{Alessandria et~al.}(2013)}]{Bolom13}
\bibinfo{author}{\bibfnamefont{F.}~\bibnamefont{Alessandria}}
  \bibnamefont{et~al.}, \bibinfo{journal}{J. Cosmol. Astropart. Phys.}
  \textbf{\bibinfo{volume}{05}}, \bibinfo{pages}{007} (\bibinfo{year}{2013}).

\bibitem[{\citenamefont{Namba}(2007)}]{ThinFoil07}
\bibinfo{author}{\bibfnamefont{T.}~\bibnamefont{Namba}},
  \bibinfo{journal}{Phys. Lett. B} \textbf{\bibinfo{volume}{645}},
  \bibinfo{pages}{398} (\bibinfo{year}{2007}).

\bibitem[{\citenamefont{Arnaboldi et~al.}(2004)}]{CUORE04}
\bibinfo{author}{\bibfnamefont{C.}~\bibnamefont{Arnaboldi}}
  \bibnamefont{et~al.} (\bibinfo{collaboration}{{The CUORE collaboration}}),
  \bibinfo{journal}{Nucl. Instrum. Meth. A} \textbf{\bibinfo{volume}{518}},
  \bibinfo{pages}{775} (\bibinfo{year}{2004}).

\bibitem[{CUO()}]{CUORE05}
\bibinfo{note}{R. Ardito et al (CUORE Collaboration), CUORE: A Cryogenic
  Underground Observatory for Rare Events, arXiv:hep-ex/0501010, (2005).}

\bibitem[{\citenamefont{Artusa et~al.}(2015)}]{CUORE17}
\bibinfo{author}{\bibfnamefont{D.~R.} \bibnamefont{Artusa}}
  \bibnamefont{et~al.} (\bibinfo{collaboration}{{The CUORE collaboration}}),
  \bibinfo{journal}{Adv. High Ener. Phys.} \textbf{\bibinfo{volume}{2015}},
  \bibinfo{pages}{879871} (\bibinfo{year}{2015}),
  \bibinfo{note}{http://dx.doi.org/10.1155/2015/870871}.

\bibitem[{Ald()}]{Alduino17}
\bibinfo{note}{C. Alduino et al., (The CUORE Collaboration) arXiv:1710.07988,
  Submitted to Phys. Rev. Lett.}

\bibitem[{\citenamefont{Li et~al.}(2016)\citenamefont{Li, Creswick, {F.T.
  Avignone III}, and Wang}}]{LiAvigWang16}
\bibinfo{author}{\bibfnamefont{D.}~\bibnamefont{Li}},
  \bibinfo{author}{\bibfnamefont{R.}~\bibnamefont{Creswick}},
  \bibinfo{author}{\bibnamefont{{F.T. Avignone III}}}, \bibnamefont{and}
  \bibinfo{author}{\bibfnamefont{Y.}~\bibnamefont{Wang}},
  \bibinfo{journal}{JCAP} \textbf{\bibinfo{volume}{02}}, \bibinfo{pages}{031}
  (\bibinfo{year}{2016}).

\bibitem[{\citenamefont{Bali et~al.}(2012)}]{QCDSF12}
\bibinfo{author}{\bibfnamefont{G.}~\bibnamefont{Bali}} \bibnamefont{et~al.},
  \bibinfo{journal}{Phys. Rev. Lett.} \textbf{\bibinfo{volume}{108}},
  \bibinfo{pages}{222001} (\bibinfo{year}{2012}).

\bibitem[{\citenamefont{Li and Thomas}(2016)}]{LiThomas15}
\bibinfo{author}{\bibfnamefont{J.}~\bibnamefont{Li}} \bibnamefont{and}
  \bibinfo{author}{\bibfnamefont{A.~W.} \bibnamefont{Thomas}},
  \bibinfo{journal}{Nuc. Phys.B} \textbf{\bibinfo{volume}{906}},
  \bibinfo{pages}{60} (\bibinfo{year}{2016}), \bibinfo{note}{arXiv:1506.03560
  [hep-ph]}.

\bibitem[{JEL()}]{JELLIS}
\bibinfo{note}{The Strange Spin of the Nucleon, J. Ellis and M. Karliner,
  hep-ph/96101280.}

\bibitem[{Mic()}]{MicroBooNE17}
\bibinfo{note}{S Pate (for the MicroBooNE Collaboration), Progress On
  Neutrino-Proton Neutral-Current Scattering In MicroBooNE,arXiv:1701.04483
  [nucl-ex]}.

\bibitem[{\citenamefont{di~Cortona et~al.}(2016)\citenamefont{di~Cortona,
  Hardy, Vega, and Villadoro}}]{CHVV16}
\bibinfo{author}{\bibfnamefont{G.~G.} \bibnamefont{di~Cortona}},
  \bibinfo{author}{\bibfnamefont{E.}~\bibnamefont{Hardy}},
  \bibinfo{author}{\bibfnamefont{J.~P.} \bibnamefont{Vega}}, \bibnamefont{and}
  \bibinfo{author}{\bibfnamefont{G.}~\bibnamefont{Villadoro}},
  \bibinfo{journal}{JHEP} \textbf{\bibinfo{volume}{01}}, \bibinfo{pages}{034}
  (\bibinfo{year}{2016}), \bibinfo{note}{arXiv:1511.02867 , (hep-ph), (hep-ex),
  (hep-lat)}.

\bibitem[{Ale({\natexlab{a}})}]{Alexandrou15}
\bibinfo{note}{Abdel-Rehim, C. Alexandrou, M. Constantinou, K. Hadjiyiannakou,
  K. Jansen, C. Kallidonis, G. Koutsou and A. V. Avils-Casco, arXiv:1511.00433
  [hep-lat].}

\bibitem[{Ale({\natexlab{b}})}]{Alexandrou16}
\bibinfo{note}{C. Alexandrou, M. Constantinou, K. Hadjiyiannakou, Ch.
  Kallidonis, G. Koutsou, K. Jansen, Ch. Wiese, A. Vaquero Avilés-Casco ,
  arXiv:1611.09163 [hep-lat].}

\bibitem[{\citenamefont{Vergados et~al.}(2016)\citenamefont{Vergados, {F. T.
  Avignone III}, Kortelainen, Pirinen, Srivastava, Suhonen, and
  Thomas}}]{Te125.16}
\bibinfo{author}{\bibfnamefont{J.}~\bibnamefont{Vergados}},
  \bibinfo{author}{\bibnamefont{{F. T. Avignone III}}},
  \bibinfo{author}{\bibfnamefont{M.}~\bibnamefont{Kortelainen}},
  \bibinfo{author}{\bibfnamefont{P.}~\bibnamefont{Pirinen}},
  \bibinfo{author}{\bibfnamefont{P.~C.} \bibnamefont{Srivastava}},
  \bibinfo{author}{\bibfnamefont{J.}~\bibnamefont{Suhonen}}, \bibnamefont{and}
  \bibinfo{author}{\bibfnamefont{A.~W.} \bibnamefont{Thomas}},
  \bibinfo{journal}{J. Phys. G: Nucl. and Part. Phys.}
  \textbf{\bibinfo{volume}{43}}, \bibinfo{pages}{11502} (\bibinfo{year}{2016}).

\bibitem[{\citenamefont{Aubert et~al.}(1983)}]{EMC83}
\bibinfo{author}{\bibfnamefont{J.~J.} \bibnamefont{Aubert}}
  \bibnamefont{et~al.}, \bibinfo{journal}{Phys. Lett. B}
  \textbf{\bibinfo{volume}{123}}, \bibinfo{pages}{275} (\bibinfo{year}{1983}).

\bibitem[{\citenamefont{Ashman et~al.}(1990)}]{EMC90}
\bibinfo{author}{\bibfnamefont{J.}~\bibnamefont{Ashman}} \bibnamefont{et~al.}
  (\bibinfo{collaboration}{{The EMC collaboration}}), \bibinfo{journal}{Nuc.
  Phys. B} \textbf{\bibinfo{volume}{328}}, \bibinfo{pages}{1}
  (\bibinfo{year}{1990}).

\bibitem[{\citenamefont{Geesaman et~al.}(1995)\citenamefont{Geesaman, Saito,
  and Thomas}}]{EMC95}
\bibinfo{author}{\bibfnamefont{D.~F.} \bibnamefont{Geesaman}},
  \bibinfo{author}{\bibfnamefont{K.}~\bibnamefont{Saito}}, \bibnamefont{and}
  \bibinfo{author}{\bibfnamefont{A.}~\bibnamefont{Thomas}},
  \bibinfo{journal}{Ann. Rev. Nuc. Sci} \textbf{\bibinfo{volume}{45}},
  \bibinfo{pages}{337} (\bibinfo{year}{1995}).

\bibitem[{\citenamefont{Cheng and Chiang}(2012)}]{ChengChiang12}
\bibinfo{author}{\bibfnamefont{H.~Y.} \bibnamefont{Cheng}} \bibnamefont{and}
  \bibinfo{author}{\bibfnamefont{C.~W.} \bibnamefont{Chiang}},
  \bibinfo{journal}{JHEP} \textbf{\bibinfo{volume}{1207}}, \bibinfo{pages}{009}
  (\bibinfo{year}{2012}), \bibinfo{note}{arXiv:1202.1292 (hep-ph)}.

\bibitem[{\citenamefont{Poves et~al.}(2001)\citenamefont{Poves,
  S\'anchez-Solano, Caurier, and Nowacki}}]{A.Poves}
\bibinfo{author}{\bibfnamefont{A.}~\bibnamefont{Poves}},
  \bibinfo{author}{\bibfnamefont{J.}~\bibnamefont{S\'anchez-Solano}},
  \bibinfo{author}{\bibfnamefont{E.}~\bibnamefont{Caurier}}, \bibnamefont{and}
  \bibinfo{author}{\bibfnamefont{F.}~\bibnamefont{Nowacki}},
  \bibinfo{journal}{Nuc. Phys. A} \textbf{\bibinfo{volume}{694}},
  \bibinfo{pages}{157} (\bibinfo{year}{2001}).

\bibitem[{\citenamefont{Kuo and Brown}(1968)}]{kbthreea}
\bibinfo{author}{\bibfnamefont{T.}~\bibnamefont{Kuo}} \bibnamefont{and}
  \bibinfo{author}{\bibfnamefont{G.}~\bibnamefont{Brown}},
  \bibinfo{journal}{Nuc. Phys. A} \textbf{\bibinfo{volume}{114}},
  \bibinfo{pages}{241} (\bibinfo{year}{1968}).

\bibitem[{\citenamefont{Poves and Zuker}(1981)}]{kbthreeb}
\bibinfo{author}{\bibfnamefont{A.}~\bibnamefont{Poves}} \bibnamefont{and}
  \bibinfo{author}{\bibfnamefont{A.}~\bibnamefont{Zuker}},
  \bibinfo{journal}{Phys. Rep.} \textbf{\bibinfo{volume}{70}},
  \bibinfo{pages}{235} (\bibinfo{year}{1981}).

\bibitem[{\citenamefont{Brown and Rae}(2014)}]{nushellx}
\bibinfo{author}{\bibfnamefont{B.}~\bibnamefont{Brown}} \bibnamefont{and}
  \bibinfo{author}{\bibfnamefont{W.}~\bibnamefont{Rae}},
  \bibinfo{journal}{Nuclear Data Sheets} \textbf{\bibinfo{volume}{120}},
  \bibinfo{pages}{115} (\bibinfo{year}{2014}).

\bibitem[{ENS()}]{ENSDF}
\bibinfo{note}{ENSDF database, http://www.nndc.bnl.gov/ensdf/.}

\bibitem[{\citenamefont{Ringwald and Saikawa}(2016)}]{RingSaika15}
\bibinfo{author}{\bibfnamefont{A.}~\bibnamefont{Ringwald}} \bibnamefont{and}
  \bibinfo{author}{\bibfnamefont{K.}~\bibnamefont{Saikawa}},
  \bibinfo{journal}{Phys. Rev. D} \textbf{\bibinfo{volume}{93}},
  \bibinfo{pages}{085031} (\bibinfo{year}{2016}),
  \bibinfo{note}{arXiv:1512.06436 [hep-ph]}.

\bibitem[{\citenamefont{Kaplan}(1985)}]{Kaplan85}
\bibinfo{author}{\bibfnamefont{D.~B.} \bibnamefont{Kaplan}},
  \bibinfo{journal}{Nuc. Phys. B} \textbf{\bibinfo{volume}{260}},
  \bibinfo{pages}{215} (\bibinfo{year}{1985}).

\bibitem[{\citenamefont{Haxton and Lee}(1991)}]{HaxLee91}
\bibinfo{author}{\bibfnamefont{W.~C.} \bibnamefont{Haxton}} \bibnamefont{and}
  \bibinfo{author}{\bibfnamefont{K.~Y.} \bibnamefont{Lee}},
  \bibinfo{journal}{Phys. Rev. Lett.} \textbf{\bibinfo{volume}{66}},
  \bibinfo{pages}{2557} (\bibinfo{year}{1991}).

\bibitem[{\citenamefont{Peccei}(1996)}]{Peccei96}
\bibinfo{author}{\bibfnamefont{R.~D.} \bibnamefont{Peccei}},
  \bibinfo{journal}{J. Korean Phys. Soc.} \textbf{\bibinfo{volume}{29}},
  \bibinfo{pages}{S199} (\bibinfo{year}{1996}), \bibinfo{note}{hep-ph/9606475}.

\bibitem[{\citenamefont{{F. T. Avignone III} et~al.}(1988)}]{Avignone88}
\bibinfo{author}{\bibnamefont{{F. T. Avignone III}}} \bibnamefont{et~al.},
  \bibinfo{journal}{Phys. Rev. D} \textbf{\bibinfo{volume}{37}},
  \bibinfo{pages}{618} (\bibinfo{year}{1988}).

\bibitem[{\citenamefont{Turck-Chiese et~al.}(1993)}]{SFluxPR93}
\bibinfo{author}{\bibfnamefont{S.}~\bibnamefont{Turck-Chiese}}
  \bibnamefont{et~al.}, \bibinfo{journal}{Phys. Rep.}
  \textbf{\bibinfo{volume}{230}}, \bibinfo{pages}{57} (\bibinfo{year}{1993}).

\bibitem[{\citenamefont{Andriamonje et~al.}(2009{\natexlab{b}})}]{SFluxCAST09}
\bibinfo{author}{\bibfnamefont{S.}~\bibnamefont{Andriamonje}}
  \bibnamefont{et~al.} (\bibinfo{collaboration}{{The CAST collaboration}}),
  \bibinfo{journal}{JCAP} \textbf{\bibinfo{volume}{0912}}, \bibinfo{pages}{002}
  (\bibinfo{year}{2009}{\natexlab{b}}), \bibinfo{note}{arXiv:0906.4488}.

\bibitem[{\citenamefont{Bahcall and A}(2006)}]{Bahcall06}
\bibinfo{author}{\bibfnamefont{J.~N.} \bibnamefont{Bahcall}} \bibnamefont{and}
  \bibinfo{author}{\bibfnamefont{M.}~\bibnamefont{A}}, \bibinfo{journal}{Astr.
  J. Suppl. Ser.} \textbf{\bibinfo{volume}{165}}, \bibinfo{pages}{400}
  (\bibinfo{year}{2006}).

\bibitem[{\citenamefont{D.B.Guenther et~al.}(1992)\citenamefont{D.B.Guenther,
  P.Dmarque, Kim, and Pinsonneault}}]{SolarT92}
\bibinfo{author}{\bibnamefont{D.B.Guenther}},
  \bibinfo{author}{\bibnamefont{P.Dmarque}},
  \bibinfo{author}{\bibfnamefont{Y.-C.} \bibnamefont{Kim}}, \bibnamefont{and}
  \bibinfo{author}{\bibfnamefont{M.~H.} \bibnamefont{Pinsonneault}},
  \bibinfo{journal}{Astr. J.} \textbf{\bibinfo{volume}{387}},
  \bibinfo{pages}{372} (\bibinfo{year}{1992}).

\bibitem[{\citenamefont{Bahcall and Pinsonneault}(1995)}]{Bahcall95}
\bibinfo{author}{\bibfnamefont{J.~N.} \bibnamefont{Bahcall}} \bibnamefont{and}
  \bibinfo{author}{\bibfnamefont{M.~H.} \bibnamefont{Pinsonneault}},
  \bibinfo{journal}{Rev. Mod. Phys.} \textbf{\bibinfo{volume}{67}},
  \bibinfo{pages}{78} (\bibinfo{year}{1995}).

\bibitem[{\citenamefont{Zioutas et~al.}(2005)}]{CAST07}
\bibinfo{author}{\bibfnamefont{K.}~\bibnamefont{Zioutas}} \bibnamefont{et~al.}
  (\bibinfo{collaboration}{{The CAST collaboration}}), \bibinfo{journal}{Phys.
  Rev. Lett.} \textbf{\bibinfo{volume}{94}}, \bibinfo{pages}{121031}
  (\bibinfo{year}{2005}), \bibinfo{note}{[hep-ex/0411033]}.

\bibitem[{\citenamefont{Anastassopoulos et~al.}(2017)}]{CAST17b}
\bibinfo{author}{\bibfnamefont{V.}~\bibnamefont{Anastassopoulos}}
  \bibnamefont{et~al.} (\bibinfo{collaboration}{{The CAST collaboration}}),
  \bibinfo{journal}{nature physics} \textbf{\bibinfo{volume}{CERN}},
  \bibinfo{pages}{584} (\bibinfo{year}{2017}).

\bibitem[{\citenamefont{Raffelt}(2007)}]{Raffelt07}
\bibinfo{author}{\bibfnamefont{G.~G.} \bibnamefont{Raffelt}},
  \bibinfo{journal}{J. Phys. A} \textbf{\bibinfo{volume}{40}},
  \bibinfo{pages}{6607} (\bibinfo{year}{2007}).

\end{thebibliography}

\end{document}